%% file: main.tex
\documentclass[conference]{IEEEtran}
\IEEEoverridecommandlockouts
\makeatletter
\renewcommand{\fnum@figure}{Fig. \thefigure}
\makeatother
\usepackage[ left=0.63in, right=0.63in, top=0.75in, bottom=1.04in]{geometry}
\usepackage[T1]{fontenc}  
\usepackage{comment}
\usepackage[noadjust]{cite}
\usepackage{filecontents}
\usepackage{graphicx}
\usepackage{epstopdf}
\usepackage{amsfonts}
\DeclareGraphicsExtensions{.pdf,.jpeg,.png,.eps}
\usepackage[cmex10]{amsmath}
\usepackage{mathabx}
\usepackage{algorithmic}
\usepackage{array}
\usepackage{mdwmath}
\usepackage{mdwtab}
\usepackage{float}
\usepackage{eqparbox}
\usepackage{url}
\usepackage{color,soul}
\usepackage{amsmath} 
\usepackage{algorithm} 
\usepackage{multirow} 

\usepackage{dirtytalk} 

\usepackage[utf8]{inputenc}
\usepackage[english]{babel}
\usepackage{fancyhdr}
\usepackage{lastpage}
\usepackage{caption}
\usepackage{subcaption}
\usepackage{adjustbox}
\usepackage{amssymb}
\usepackage{amsfonts}

\usepackage{xcolor}

\usepackage{calrsfs}
\DeclareMathAlphabet{\pazocal}{OMS}{zplm}{m}{n}

\input{acronyms}
\def\BibTeX{{\rm B\kern-.05em{\sc i\kern-.025em b}\kern-.08em
    T\kern-.1667em\lower.7ex\hbox{E}\kern-.125emX}}

\begin{document}

\bstctlcite{IEEEexample:BSTcontrol}
\title{PrivFly: A Privacy-Preserving Self-Supervised Framework for Rare Attack Detection in IoFT
}

\author{\IEEEauthorblockN{Safaa Menssouri, and El Mehdi Amhoud
}
\IEEEauthorblockA{College of computing, Mohammed VI Polytechnic University, Ben Guerir, Morocco 
} 
{\{safaa.menssouri, elmehdi.amhoud\}@um6p.ma}}

\maketitle

\begin{abstract}
The Internet of Flying Things (IoFT) plays a vital role in modern applications such as aerial surveillance and smart mobility. However, it remains highly vulnerable to cyberattacks that threaten the confidentiality, integrity, and availability of sensitive data. Developing effective intrusion detection systems (IDS) for IoFT networks faces key challenges, including data imbalance, privacy concerns, and the limited capability of traditional models to detect rare but potentially damaging cyber threats. In this work, we propose PrivFly, a privacy-preserving IDS framework that integrates self-supervised representation learning and differential privacy (DP) to enhance detection performance in imbalanced IoFT network traffic. We propose a masked feature reconstruction module for self-supervised pretraining, improving feature representations and boosting rare-class detection. Differential privacy is applied during training to protect sensitive information without significantly compromising model performance. In addition, we conduct a SHapley additive explanations (SHAP)-based analysis to evaluate the impact of DP on feature importance and model behavior. Experimental results on the ECU-IoFT dataset show that PrivFly achieves up to 98\% accuracy and 99\% F1-score, effectively balancing privacy and detection performance for secure IoFT systems.
\end{abstract}

\section{Introduction And Related Works}
The \gls{IoFT} \cite{ZAIDI202153} is an emerging paradigm that integrates aerial devices such as drones and \gls{UAVs} into intelligent, interconnected ecosystems. These systems are increasingly deployed in critical domains, including military surveillance, disaster response, medical and goods delivery, and emergency aid \cite{Applications}. With the integration of IoFT into next-generation 6G infrastructures, these networks are expected to provide ultra-low latency and high reliability, enabling time-sensitive services. However, the openness of wireless communication channels and the mobility of aerial platforms significantly increase their vulnerability to cyber threats such as jamming, spoofing, and data manipulation.

To mitigate these risks, \gls{IDS} play a crucial role in securing IoFT networks. In particular, anomaly-based IDS (AIDS) \cite{choudhary2018intrusion} have proven effective in detecting both known and unknown attacks. Despite their potential, developing robust \gls{IDS} solutions for IoFT remains challenging due to several factors: (1) the lack of real-world IoFT datasets encompassing diverse attack types; (2) severe class imbalance, where rare but critical threats are underrepresented; and (3) the absence of privacy-preserving mechanisms, despite the sensitive nature of aerial traffic data.

Recent research has explored various approaches to improve intrusion detection and preserve privacy in \gls{IoT} and \gls{IoFT} environments. For example, Fahad et al. proposed a differentially private anomaly detection framework that aims to balance utility and privacy \cite{Fahad_2024}. Although their approach offers strong privacy guarantees, it is designed for general IoT scenarios and does not directly address the unique challenges posed by aerial networks. In another \mbox{study \cite{s23198077}}, a privacy-preserving IDS was developed for UAV networks using lightweight cryptography and \gls{ML} models. While promising, this work lacks support for multiclass classification and does not explicitly handle rare attacks common in IoFT environments. Similarly, the authors of \cite{10073944} focused on Sybil attack detection for FANETs-based IoFT using physical-layer characteristics. Their ML-based scheme achieved over 91\% classification accuracy; however, it was limited to Sybil attacks and lacked evaluation across diverse threat types. Other works such as \cite{vuong2024} have combined autoencoder-based feature extraction with ML models for UAV anomaly detection, yet their approaches lack support for multiclass scenarios and overlook privacy concerns. More recently, the authors of \cite{10947471} designed a generative adversarial network (GAN)-based IDS for IoFT using the ECU-IoFT dataset. They enhanced model robustness through adversarial training, and their results showed that the Random Forest (RF) model achieved the highest detection accuracy, up to 96.5\%. However, their approach was limited to binary classification, reducing its applicability in more complex, real-world scenarios. \\
Overall, while existing studies provide valuable insights, they fall short of delivering comprehensive, interpretable, and privacy-preserving IDS solutions capable of handling class imbalance and rare threats in IoFT environments.

To bridge this gap, we propose PrivFly, a privacy-preserving IDS framework that integrates self-supervised representation learning and differential privacy, enabling effective anomaly detection even with scarce and imbalanced datasets. The main contributions of this paper are summarized as follows:
\begin{itemize}
    \item \textbf{PrivFly Framework:} We propose a novel intrusion detection framework specifically designed for IoFT. To the best of our knowledge, this is the first work to integrate self-supervised learning and differential privacy into a unified pipeline for detecting rare attacks in IoFT environments.
    \item  \textbf{Improved Rare Attack Detection on ECU-IoFT data:} We introduce two key components: (1) synthetic oversampling and (2) a self-supervised representation learning module that significantly improves model generalization and rare attack detection under severe class imbalance.
    \item \textbf{Explainability via SHapley Additive Explanations (SHAP) Analysis:} We conduct a SHAP-based explainability analysis to understand how individual features influence model predictions, and we evaluate the impact of differential privacy on feature importance.
\end{itemize}
 \begin{figure}[t]
\centering
\includegraphics[width=0.489\textwidth]{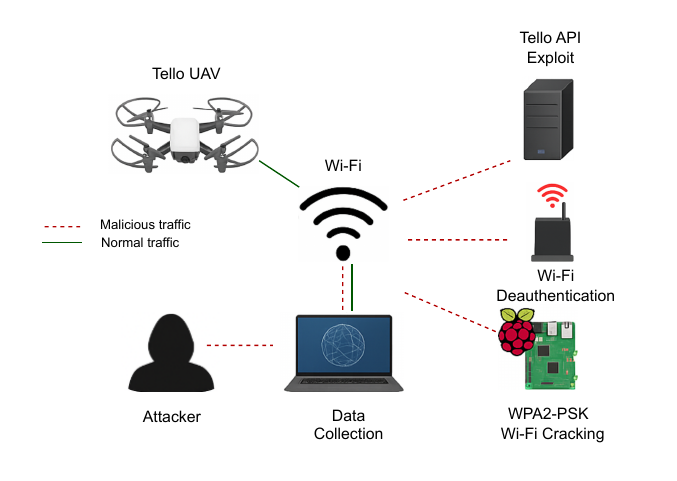}
    \caption{System model.}
    \label{fig: System model}
\end{figure}
The remainder of the paper is organized as follows: \\ In section II, we detail the architecture of our proposed PrivFly system along with a description of the dataset. In section III, we present our simulation results, and we discuss the findings. Finally, in section IV, we conclude and outline our perspectives.
\section{System Model And Methodology}
Figure \ref{fig: System model} illustrates our system model. A Ryze Tello UAV is connected to a Wi-Fi network that carries both benign and malicious traffic. A legitimate ground station controls the UAV and captures surrounding network traffic for analysis. Meanwhile, an attacker attempts to compromise the UAV using three distinct attack vectors: (1) Wi-Fi deauthentication, forcing the UAV to disconnect from its controller; (2) WPA2-PSK cracking, capturing the four-way handshake and performing an offline dictionary attack to recover the pre-shared key; and (3) Tello API exploit, using recovered credentials to issue critical commands due to the API’s lack of authentication. This setup enables comprehensive traffic data collection.

\subsection{ECU-IoFT Dataset}
In our study, we employed the open-source ECU-IOFT
dataset \cite{app12041990}, which was developed to support cybersecurity research in UAV-based IoT environments. It captures real-world traffic generated by a Tello UAV communicating over Wi-Fi in a controlled environment. Notably, the dataset exhibits a severe class imbalance, with the Tello API Exploit attack accounting for only 0.05\% of the total number of traffic instances. This rarity poses a significant challenge for conventional IDS, underscoring the need for more effective detection strategies.

\subsection{SMOTE Method}
The synthetic minority over-sampling technique (SMOTE) is a data augmentation method used to address class imbalance in ML tasks \cite{10.5555/1622407.1622416}. Rather than simply duplicating minority class samples which can lead to overfitting, SMOTE generates synthetic examples by interpolating between existing minority instances and their nearest neighbors. This process results in more diverse and representative training data, enabling classifiers to better learn the decision boundaries for minority classes.
\subsection{Conditional Tabular GAN (CTGAN)}
CTGAN is a GAN-based model tailored for generating realistic tabular data \cite{NEURIPS2019_254ed7d2}. It effectively handles mixed data types and complex feature dependencies through adversarial training. 
CTGAN employs mode-specific normalization using a Gaussian mixture model for continuous variables and one-hot encoding for categorical variables. The overall probability density function $ p(x) $ of a continuous variable $ x $ is given by: \(
p(x) = \sum_{k=1}^{K} \pi_k \mathcal{N}(x \mid \mu_k, \sigma_k^2),\)
where \( p(x) \) is formed by a weighted sum of \( K \) Gaussian components. Each component is characterized by its mean \( \mu_k \), variance \( \sigma_k^2 \), and weight \( \pi_k \) \cite{11162182}.
\begin{figure*}[t]
\centering
\includegraphics[width=0.99\textwidth]{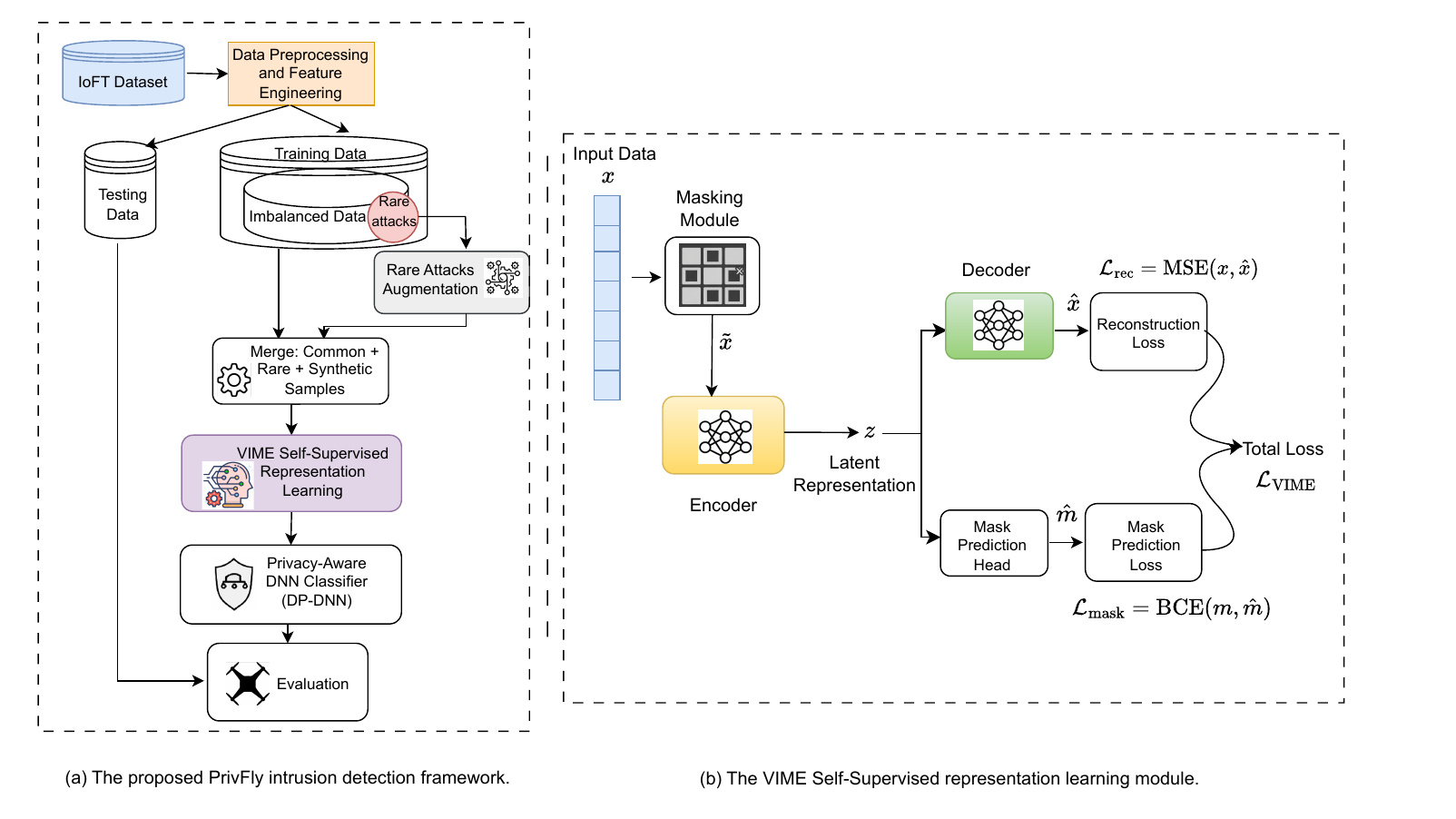}
\vspace{-0.3em}
    \caption{The architecture of the proposed PrivFly intrusion detection framework along with the VIME self-supervised representation learning module.}
    \label{fig: PrivFly IDS}
    \vspace{-0.4em}
\end{figure*}

\subsection{Problem Setup}
Let \( \mathcal{D} = \{(x_i, y_i)\}_{i=1}^N \) denote a labeled tabular dataset, where \( x_i \in \mathbb{R}^d \) is the feature vector of the $i$-th sample and \( y_i \in \{1, \ldots, K\} \) is the corresponding class label. The dataset is highly imbalanced. To address this, we augment rare classes with synthetic samples prior to training.\\ Our goal is to train an effective, privacy-preserving classifier capable of generalizing across both majority and rare classes.

\subsubsection{Self-Supervised Pretraining with VIME}
To improve feature learning prior to classification, we propose a self-supervised pretraining strategy inspired by VIME (value imputation and mask estimation) \cite{NEURIPS2020_7d97667a}. Unlike the original VIME formulation, which replaces masked features with values sampled from their marginal distributions, our approach simplifies the pretext task by applying zero-masking only. This design choice not only reduces computational complexity but also avoids introducing synthetic distributional noise, aligning more naturally with our privacy-preserving setting, as it does not rely on global data statistics for imputation. Furthermore, our framework decouples VIME's self-supervised pretraining from its semi-supervised fine-tuning component, using the learned representations as input to a separate, privacy-aware classifier.

For each input sample \( x \in \mathbb{R}^d \), we generate a binary mask vector \( m \in \{0,1\}^d \), where each component \( m_j \) is sampled independently from \(\mathrm{Bernoulli}(1-p)\), with \( p \) denoting the masking probability. The masked input is computed as:
\( \tilde{x} = m \odot x ,\)
where \( \odot \) denotes element-wise multiplication. 

The corrupted input \( \tilde{x} \) is then passed through an encoder network to produce a latent representation \( z \). Two auxiliary heads are trained jointly with the encoder: (1) a \textbf{decoder} that reconstructs the original input, with the output denoted as \( \hat{x} \), and (2) a \textbf{mask predictor} that estimates the binary mask, denoted as \( \hat{m} \). The training objective combines reconstruction loss and mask prediction loss as follows:
\[ \mathcal{L}_{\text{VIME}} =\alpha \cdot {\frac{1}{d} \sum_{j=1}^{d} \left(x_j - \hat{x}_j\right)^2} +  \text{BCE}(m, \hat{m}),\]
where \( \alpha \) is a weighting parameter, and \( \text{BCE}(m, \hat{m}) \) denotes the binary cross-entropy loss. \\After pretraining, the decoder and mask predictor are discarded. The encoder is used to extract latent representations \( z = f_{\text{enc}}(x) \), which serve as input to the downstream classifier.
\subsubsection{Differential Privacy}
To ensure privacy, we employ differentially private Stochastic Gradient Descent (DP-SGD) with gradient clipping and Gaussian noise injection \cite{ACM}. Each per-sample gradient \( g_i \) is clipped to an \(L_2\) norm threshold \( C \):
\[\tilde{g}_i = g_i \cdot \min\left(1, \frac{C}{\|g_i\|_2}\right).\]
The final gradient update, with added noise, is given by:\[
\bar{g} = \frac{1}{B} \left( \sum_{i=1}^B \tilde{g}_i + \mathcal{N}(0, \sigma^2 C^2 \mathbf{I}) \right).
\]
Here, $B$ is the batch size and \(\sigma\) controls the privacy–utility trade-off. This mechanism provides \((\varepsilon, \delta)\)-differential privacy, where $\delta > 0$ denotes the acceptable probability of privacy leakage, and $\varepsilon$ quantifies the worst-case privacy loss over multiple training steps.
\begin{table*}[ht]
\centering
\caption{Comparison of models' performance under differential privacy with varying noise multipliers}
\label{tab:combined_results}
\begin{tabular}{|c| c| c c c c c |}
\hline
\textbf{Noise Multiplier}  & \textbf{Privacy Budget} & \textbf{Method} & \textbf{Accuracy} & \textbf{Precision} & \textbf{Recall} & \textbf{F1-score} 
\\[0.3em] \hline \hline  
   &  & DP-DNN & 93 & 67 & 71 & 69 \\
   5 & 0.18 & PrivFly + CTGAN & 95 & 94 & 91 & 91 \\
    & & PrivFly + SMOTE & 95 & 94 & 91 & 92 \\
\hline 
  &  & DP-DNN & 95 & 69 & 72 & 70  \\ 
  3 & 0.5 & PrivFly + CTGAN & 96 & 97 & 92 & 94 
    \\ 
    & & PrivFly + SMOTE & 97 & 97 & 92 & 94 
    \\ 
\hline
  &  & DP-DNN & 96 & 70 & 73 & 71 \\
  1 & 1.41 & PrivFly + CTGAN & 97 & 98 & 92 & 95 \\
  &  & PrivFly + SMOTE & 98 & 98 & 93 & 95 \\
\hline
  &  & DP-DNN & 97 & 73 & 74 & 73 \\
  0.5 & 11.17 & PrivFly + CTGAN & 98 & 99 & 92 & 95 \\
    & & PrivFly + SMOTE & 98 & 99 & 99 & 99 \\
\hline
   & & DP-DNN & 98 & 73 & 74 & 74 \\
 0.2 &  219.11 & PrivFly + CTGAN & 98 & 99 & 92 & 95 \\
   &  & PrivFly + SMOTE & 98 & 99 & 99 & 99 \\
\hline
\end{tabular}
\end{table*}

\subsection{PrivFly IDS framework}
The proposed PrivFly intrusion detection framework is designed to detect rare attacks in \gls{IoFT} environments while maintaining privacy during model training. The system comprises a sequence of key stages, as illustrated in Fig. 2a. The process begins by importing the ECU-IoFT dataset, which supports the multiclass classification task. Subsequently, the dataset undergoes essential preprocessing steps such as normalization, and feature engineering to prepare the data for further analysis. 

To address the rarity of certain attacks, we performed a rare attack augmentation step using CTGAN or SMOTE methods, increasing the minority attack class from 10 to 500 total instances. This target size was chosen to ensure sufficient representation for model training while avoiding excessive synthetic oversampling that could lead to overfitting. Afterward, the generated synthetic samples are merged with both common and rare classes to produce a more enriched training set. In the next phase, we apply the VIME self-supervised learning module as depicted in Fig. 2b, which enhances feature representations by reconstructing masked features, thereby learning robust, context-aware embeddings from tabular data. The resulting representations are then passed to a differentially private deep neural network (DP-DNN) classifier, where differential privacy is enforced during the training stage of the model to protect sensitive information. We employed the standard cross-entropy loss to train our model. Finally, we tested PrivFly on the unseen 30\% dataset using various metrics, providing a thorough evaluation of the model’s capability to detect rare classes.
\subsection{Performance Evaluation Metrics}
To assess the performance of our proposed \gls{IDS}, we used various metrics, including accuracy ($AC$), precision ($PR$), recall ($R$), and F1-score ($F1$) via confusion matrix. These performance metrics are defined as follows:
\begin{equation}
AC=\frac{TP+TN}{TP+TN+FP+FN}, \mathrm{~}  R=\frac{TP}{TP+FN},
\end{equation}
\begin{equation}
F1=\frac{2TP}{2TP+FP+FN}, \mathrm{~}  PR=\frac{TP}{TP+FP}.
\end{equation}

With $TP$, $TN$, $FP$ and $FN$ representing true positives, true negatives, false positives, and false negatives, respectively.

\begin{figure}[t] 
\centering
\includegraphics[scale=0.63]{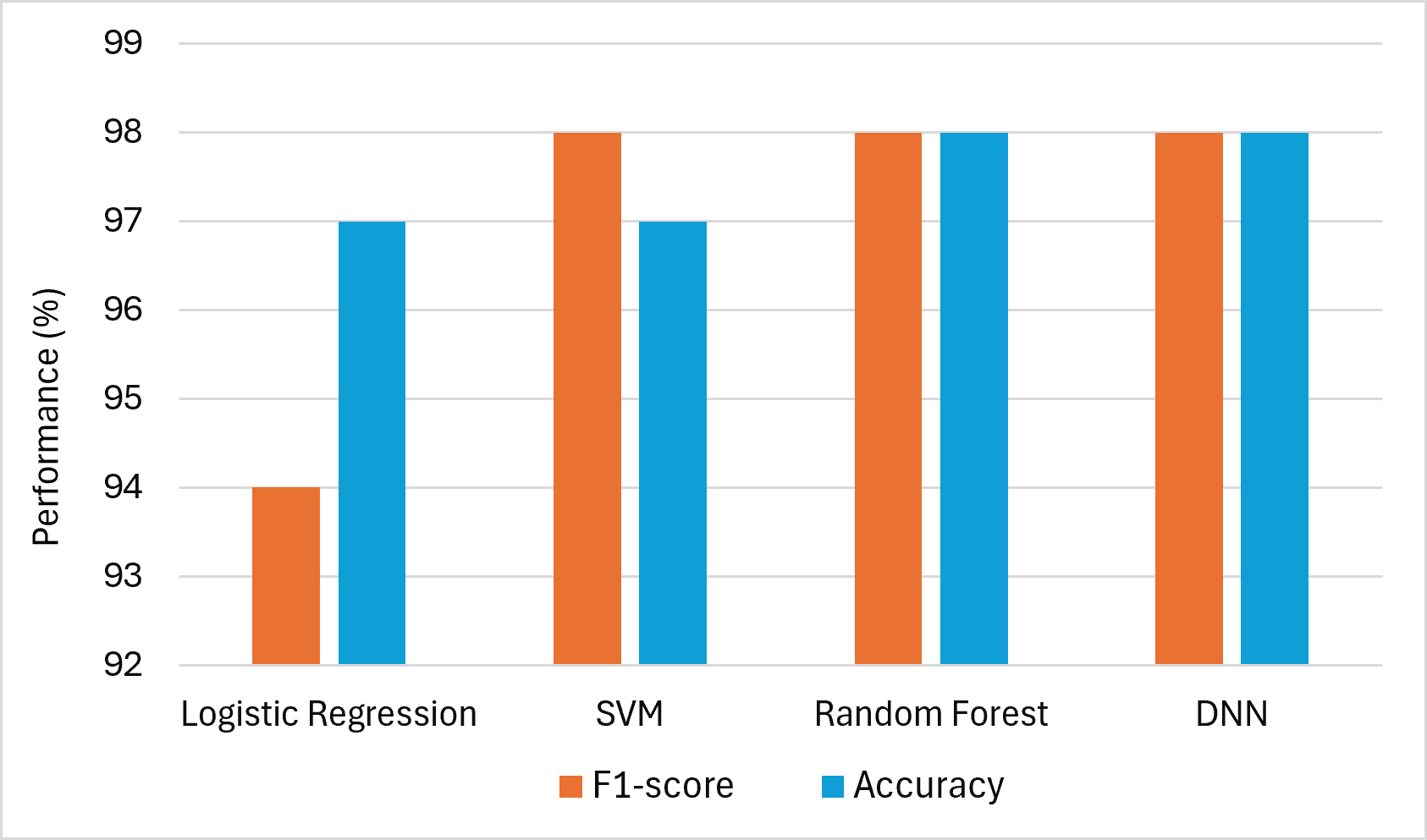}
     \caption{Performance comparison of classifiers based on Accuracy and F1-Score.}
        \label{fig: ComparaisonACCModels}
\end{figure}

\section{Experimental Results}
This section presents the performance results of our framework, along with SHAP-based analysis for the selected models.

\subsection{Baseline Comparison and Model Selection}
To determine the most suitable baseline for our study, we evaluated several classifiers on the unseen data after training them using a five-fold cross validation strategy: Logistic Regression, Support Vector Machine (SVM), Random Forest, and DNN on the original, non-private dataset. As illustrated in \mbox{Fig. \ref{fig: ComparaisonACCModels}}, both RF and DNN models achieved high F1-scores and accuracy up to 98\%. However, the DNN is inherently more compatible with DP mechanisms such as DP-SGD, which are crucial to our framework. Therefore, we selected the DNN as the core model for all subsequent experiments.
\begin{figure*}[t]
\centering
    \begin{subfigure}[b]{0.4\textwidth}
        \includegraphics[width=\textwidth]{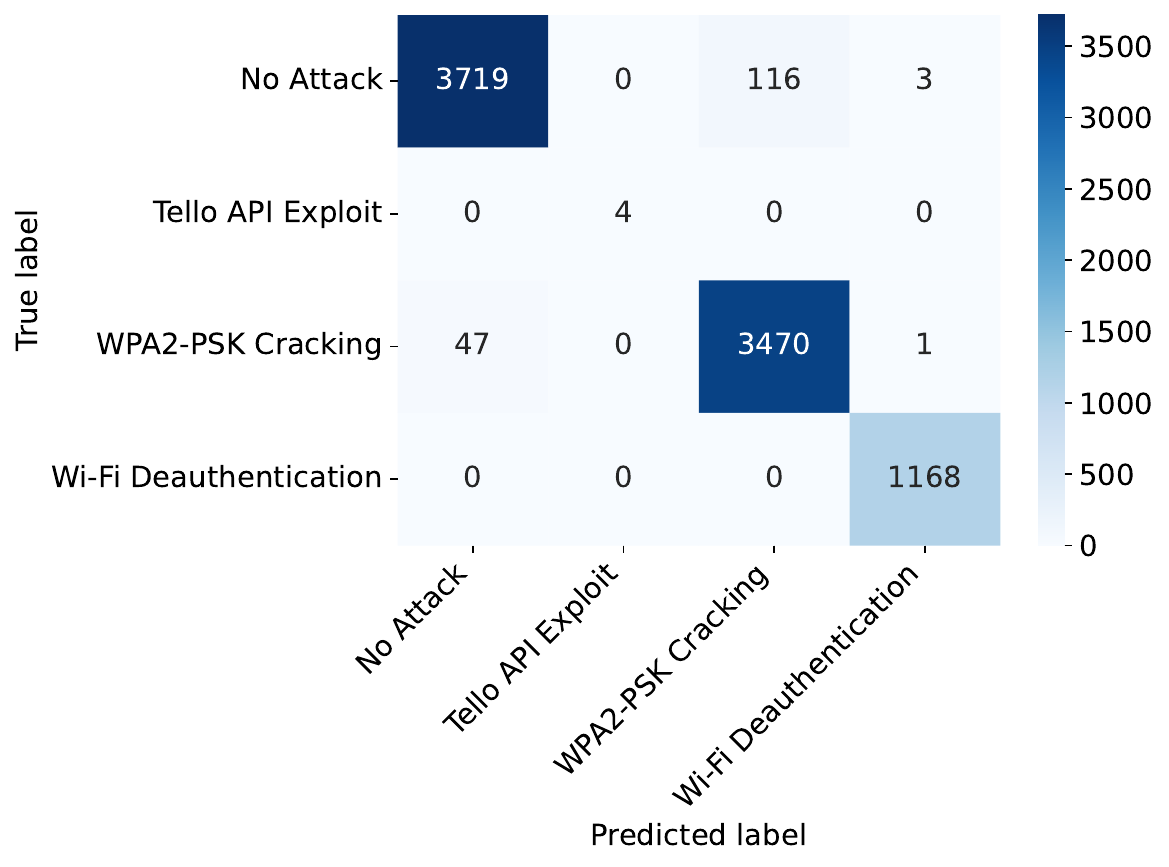}
        \caption{Confusion matrix for DNN-based multiclass classifier.}
        \label{fig:ConfusionMatrixDNN}
    \end{subfigure}
    \begin{subfigure}[b]{0.56\textwidth}
        \includegraphics[width=\textwidth]{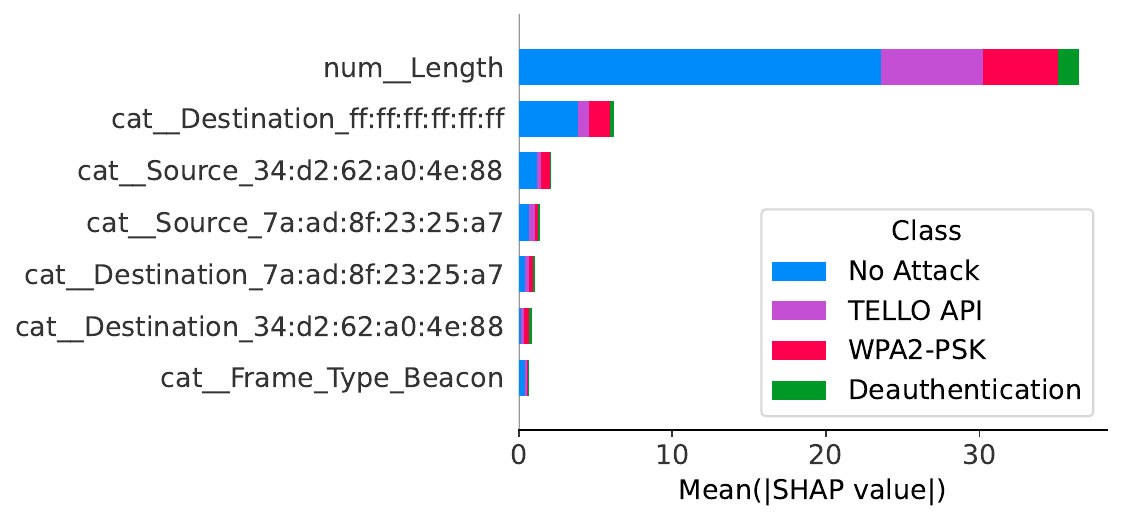}
        \caption{SHAP summary plot for the DNN model without DP.}
        \label{fig:SHAPDNN}
    \end{subfigure}
    \begin{subfigure}[b]{0.39\textwidth}
    \vspace{1em}
        \includegraphics[width=\textwidth]{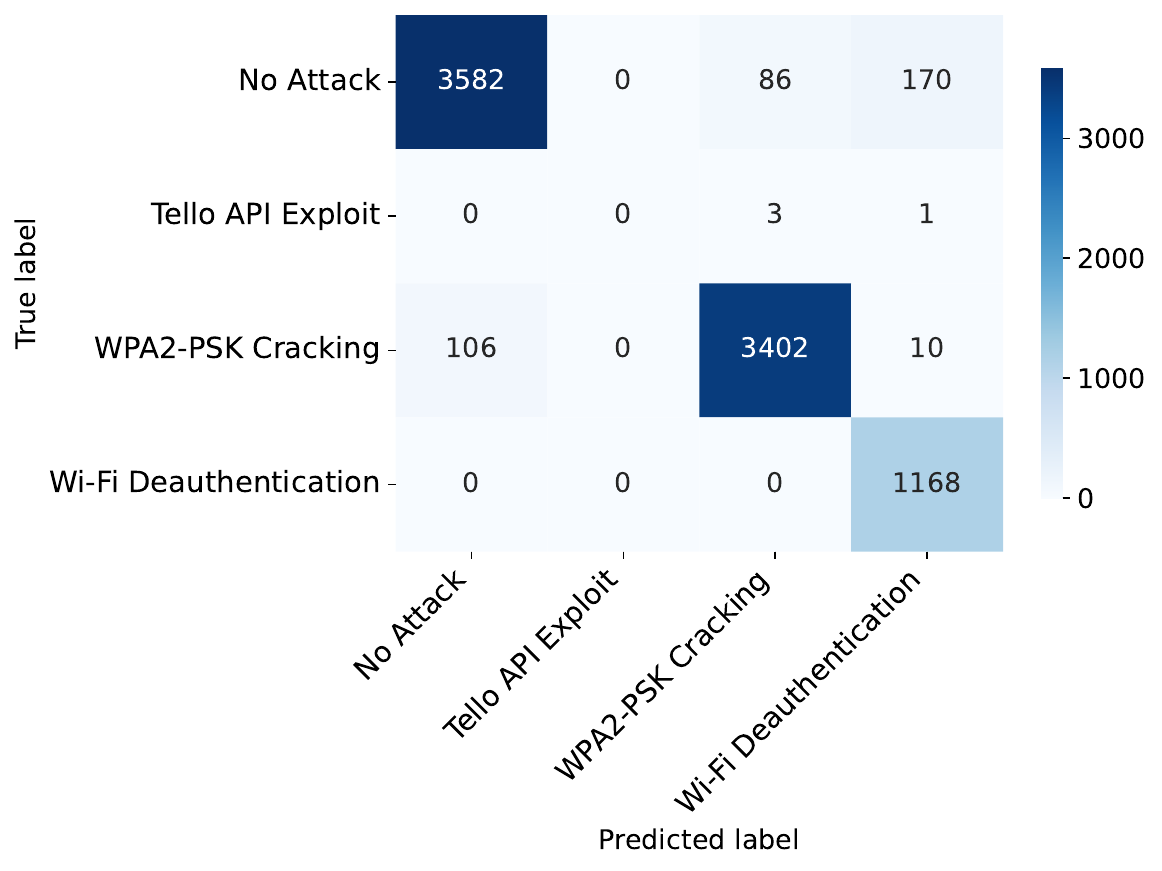}
        \caption{Confusion matrix for the DNN-based multiclass classifier under DP.}
        \label{fig:ConfusionDNNDP}
    \end{subfigure}
    \begin{subfigure}[b]{0.56\textwidth}
        \includegraphics[width=\textwidth]{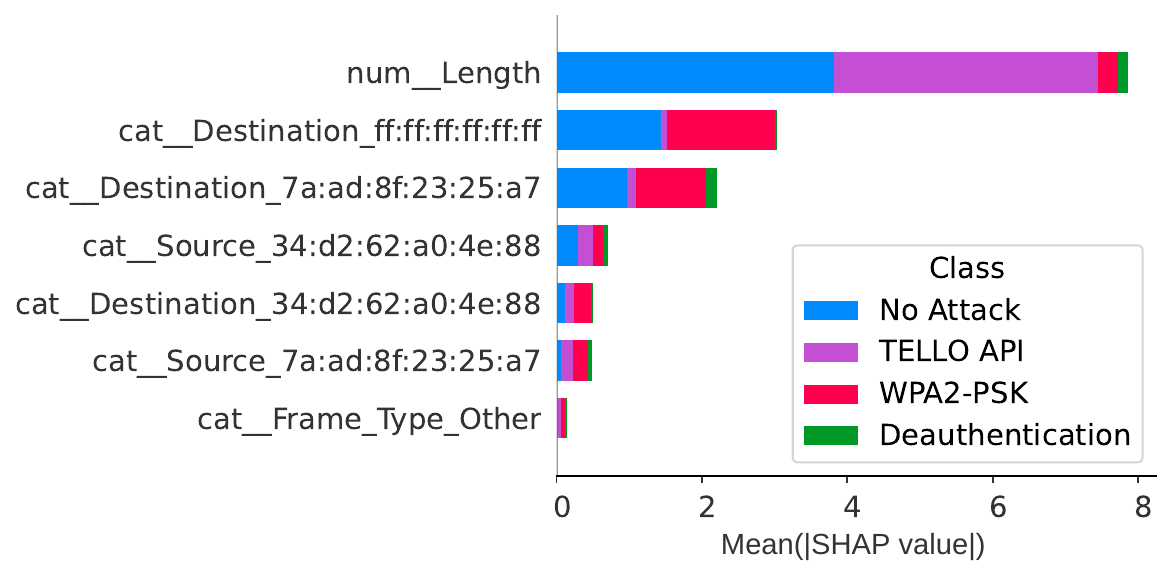}
        \caption{SHAP summary plot for the DNN model under DP.}
        \label{fig:SHAPDNNDP}
    \end{subfigure}

    \caption{Performance evaluation and SHAP-based explainability analysis of the DNN model with and without differential privacy (DP).}
    \label{fig:ConfusionAndSHAP}
\end{figure*}

\subsection{Performance and Explainability Analysis of the Model}
To further evaluate the performance of the selected DNN model, we analyzed its behavior on the original, non-private dataset. As shown in Fig. \ref{fig:ConfusionMatrixDNN}, the multi-class classifier achieves strong performance across all classes on unseen data (30\%), including minority categories such as Tello API Exploit, despite the pronounced class imbalance. This suggests that the DNN effectively captures meaningful patterns from raw network traffic data without requiring explicit balancing techniques.

To gain insight into the decision-making process of the model, we conducted a SHAP-based explainability analysis. From Fig. \ref{fig:SHAPDNN} we observe that key features such as frame length, broadcast destination addresses, and specific MAC identifiers (e.g., 7a:ad:8f:23:25:a7) have the highest mean SHAP values, indicating their significant global impact on the model's predictions. These features align with known traffic characteristics and appear to provide reliable decision boundaries across all classes. Interestingly, while not visible in the global ranking, further analysis revealed that the source IP 192.168.10.1, a rare but class-specific indicator, is used in conjunction with core features like packet length to enhance Tello API classification. This suggests that the model combines both general traffic patterns and environment-specific signatures to accurately classify rare attacks. However, this behavior also raises concerns about overfitting. The model’s apparent reliance on fixed device identifiers and static IP addresses, particularly for rare classes, may reduce its robustness and generalizability to more dynamic or adversarial environments. These observations motivated us to introduce differential privacy in subsequent experiments to regularize feature attributions and improve generalization.

\subsection{Evaluation of DP Integration and Framework Performance}
\begin{figure}[t] 
\centering
\includegraphics[scale=0.4]{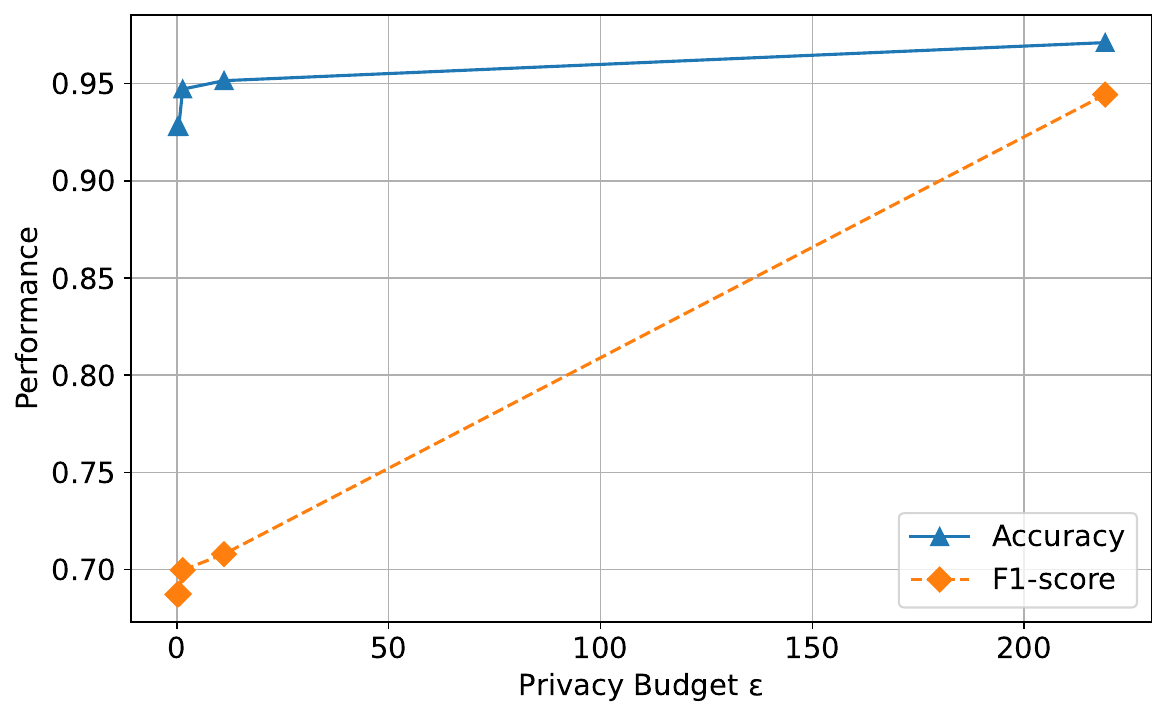}
        \caption{Trade-Off between privacy budget and model performance.}
        \label{fig: TradeOff}
\end{figure}

Introducing differential privacy to the DNN model significantly altered its behavior. As shown in Fig. \ref{fig:ConfusionDNNDP}, applying DP with a moderate noise multiplier (0.5) leads to reduced performance, particularly in detecting rare classes such as Tello API. The classifier maintains good performance on majority classes, but completely fails to detect the rare class Tello (0 from 4). This drop in performance detection is attributed to the noise introduced by DP, which disrupts the model’s ability to capture sharp, class-specific patterns associated with underrepresented attacks. This trend is further confirmed by the SHAP summary plot in Fig. \ref{fig:SHAPDNNDP}. Compared to the non-private model, the SHAP values for rare-class specific features such as broadcast and device identifiers are significantly diminished, as reflected by the lower mean SHAP values under DP. The model relies more heavily on general features like packet length, which may still provide adequate separation for common classes but are insufficient to detect rare attacks. This shows that DP improves privacy at the cost of model sensitivity to minority patterns.\\
To better understand the privacy-utility trade-off, we varied the noise multiplier and evaluated the corresponding F1-scores and accuracy metrics, as shown in Table \ref{tab:combined_results} and Fig. \ref{fig: TradeOff}. Our findings confirm the expected trade-off: stronger privacy (i.e., lower $\varepsilon$) leads to a decrease in model performance. This is because a higher noise multiplier introduces more randomness to protect individual data points, which in turn reduces the model's ability to learn precise patterns. Consequently, for applications with high data sensitivity, such as military surveillance, a strict privacy budget like $\varepsilon = 0.18$ (noise multiplier of 5) would be necessary to ensure maximum protection, even at the cost of a reduced F1-score. Conversely, for applications with lower privacy requirements, such as monitoring public air traffic, a higher privacy budget like $\varepsilon = 11.17$ (noise multiplier of 0.5) could offer a good balance between privacy and performance.

\begin{figure}[t]  
\centering
\includegraphics[scale=0.39]{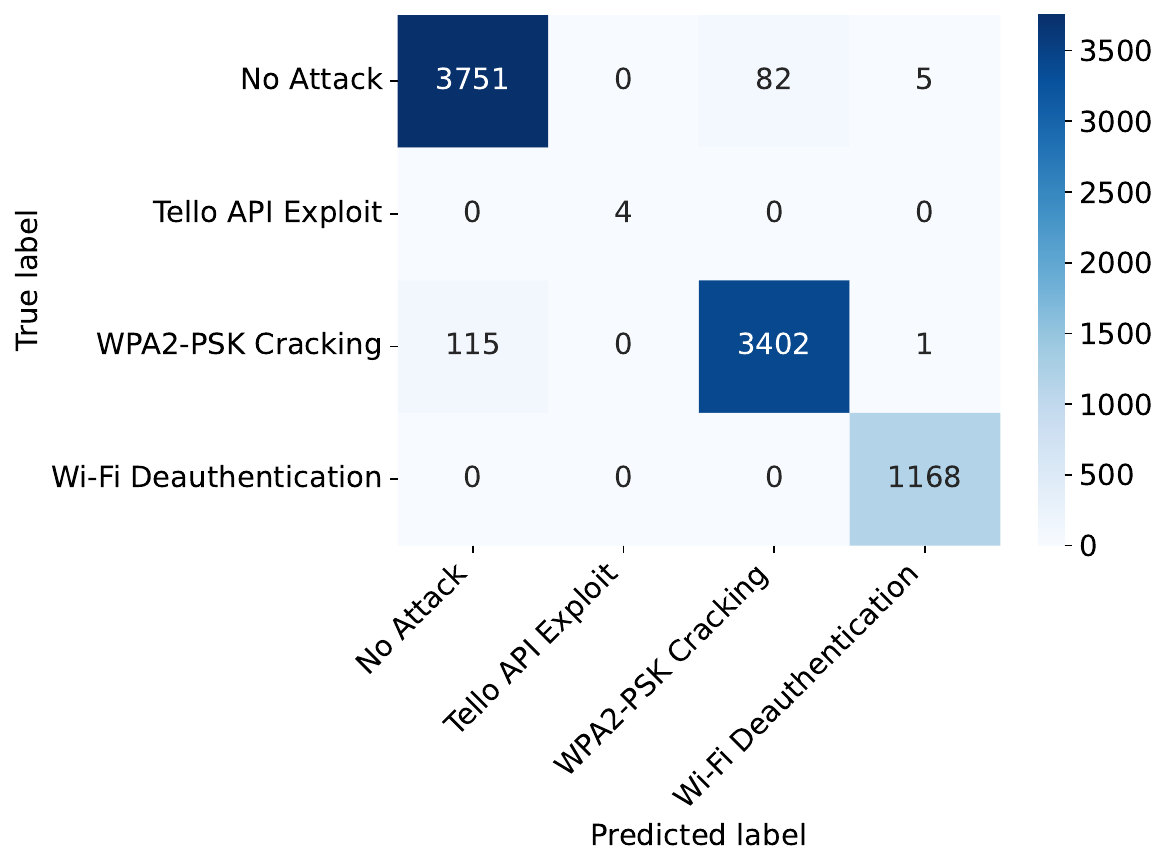}
\vspace{-0.5em}
     \caption{Confusion matrix of the proposed PrivFly framework.}
        \label{fig: ConfusionPrivFly}
\end{figure}

To address the challenge of maintaining effective intrusion detection while enforcing privacy guarantees, we introduced two key enhancements: (1) synthetic oversampling to mitigate rare-class sparsity, and (2) a self-supervised representation learning module to improve the model’s generalization. As illustrated in Tables \ref{tab: EvaluationNoise0} and \ref{tab: EvaluationNoise3} respectively, these enhancements significantly recover the performance lost due to DP. For example, at a noise multiplier of 0.5, the F1-score improves from 73\% (baseline DP-DNN) to 99\% with PrivFly, enhancing robustness without compromising accuracy. At a more constrained setting with noise equal to 3, the F1-score improves from 70\% to 94\%, demonstrating the effectiveness of PrivFly.\\
Building on our previous analysis, Table \ref{tab:combined_results} shows that the proposed PrivFly framework achieves substantial improvements across all noise levels compared to the baseline DP-DNN model. It reaches up to 92\% F1-score under high noise settings, and achieves up to 98\% for all metrics performance at moderate privacy levels effectively mitigating the performance degradation typically caused by DP. The confusion matrix in Fig. \ref{fig: ConfusionPrivFly} further confirms that the rare classe Tello API, previously undetected by the baseline model, is correctly classified. These results confirm the efficacy of our framework.

\begin{table}[t]
 \centering
  \caption{Macro-averaged evaluation of models under DP (noise = 0.5)}
    \centering
    \begin{tabular}{|p{2.5cm}|c|c|c|c|}
    \hline
    \centering
    \textbf{Model} & \textbf{Accuracy} & \textbf{Precision} &  \textbf{Recall} &  \textbf{F1-score}\\
       \hline
       \centering
        DP-DNN & 97 & 73 & 74 & 73  \\
        \hline
        \centering
        DP-DNN (focal loss) & 97 & 73 & 74 & 73 \\ 
       \hline
        \centering
        DP-DNN (+ class weights) & 97 & 73 & 74 & 73 \\ 
        \hline
        \centering
        DP-DNN + VIME & 98 & 74 & 74 & 74 \\ 
       \hline
        \centering
        DP-DNN + CTGAN & 97 & 98 & 92 & 95\\ 
       \hline
       \centering
        DP-DNN + SMOTE & 97 & 98 & 92 & 95 \\
        \hline
        \centering
       PrivFly (CTGAN+VIME)& 98 & 99 & 92 & 95 \\
       \hline
       \centering
       \textbf{PrivFly (SMOTE+VIME)}& \textbf{98} & \textbf{99} & \textbf{99} & \textbf{99}\\
       \hline     
    \end{tabular}
    \label{tab: EvaluationNoise0}
    \vspace{-0.5em}
\end{table}

\begin{table}[t]
 \centering
  \caption{Macro-averaged evaluation of methods under DP (noise = 3)}
    \centering
    \begin{tabular}{|p{2.5cm}|c|c|c|c|}
    \hline
    \centering
    \textbf{Model} & \textbf{Accuracy} & \textbf{Precision} &  \textbf{Recall} &  \textbf{F1-score}\\
       \hline
       \centering
        DP-DNN & 95 & 69 & 72 & 70  \\
        \hline
        \centering
        DP-DNN (focal loss) & 95 & 69 & 72 & 70 \\ 
       \hline
        \centering
        DP-DNN (+ class weights) & 95 & 69 & 72 & 70 \\ 
        \hline
        \centering
        DP-DNN + VIME & 97 & 72 & 73 & 73 \\ 
       \hline
        \centering
        DP-DNN + CTGAN & 95 & 94 & 91 & 91\\ 
       \hline
       \centering
        DP-DNN + SMOTE & 95 & 94 & 91 & 91 \\
        \hline
        \centering
       PrivFly (CTGAN+VIME)& 96 & 97 & 92 & 94 \\
       \hline
       \centering
       \textbf{PrivFly (SMOTE+VIME)}& \textbf{97} & \textbf{97} & \textbf{92} & \textbf{94}\\
       \hline      
    \end{tabular}
    \label{tab: EvaluationNoise3}
\end{table}

\section{Conclusion}
In this paper, we proposed PrivFly, a novel privacy-aware intrusion detection framework tailored for securing IoFT networks. By integrating self-supervised representation learning with differential privacy, PrivFly effectively addresses the challenges of rare attack detection and privacy-preserving training. Experimental evaluation on the ECU-IoFT dataset demonstrates that PrivFly achieves up to 99\% F1-score at moderate privacy levels, while maintaining a robust 92\% F1-score under strict privacy constraints. To enhance model transparency, we conducted a SHAP-based explainability analysis, identifying key influential features and providing valuable insights into the model's decision-making process. In our future work, we plan to extend PrivFly to distributed learning scenarios and assess its robustness against adversarial threats, thereby improving its practical deployment in real-world IoFT environments.
\section*{Acknowledgment}
\setlength{\parindent}{0cm}
This work was sponsored by the 100 PhDs for Africa programme under the UM6P-EPFL Excellence in Africa Initiative.

\bibliographystyle{IEEEtran}
\bibliography{biblio_traps_dynamics}
\end{document}

%% file: acronyms.tex
\usepackage[acronym]{glossaries}
\newacronym{IDS}{IDS}{intrusion detection systems}
\newacronym{CNN}{CNN}{convolutional neural networks}
\newacronym{LSTM}{LSTM}{long short-term memory}
\newacronym{AI}{AI}{artificial intelligence}
\newacronym{DL}{DL}{deep learning}
\newacronym{ML}{ML}{machine learning}
\newacronym{FP}{FP}{false positives}
\newacronym{FN}{FN}{false negatives}
\newacronym{TP}{TP}{true positives}
\newacronym{TN}{TN}{true negatives}
\newacronym{UAVs}{UAVs}{unmanned aerial vehicles}
\newacronym{UAV-IDS}{UAV-IDS}{unmanned aerial vehicle intrusion detection system}
\newacronym{PCA}{PCA} {principal component analysis}
\newacronym{Relu}{Relu}{rectified linear unit}
\newacronym{IoT}{IoT}{Internet of Things}
\newacronym{SIDS}{SIDS}{signature-based IDS}
\newacronym{AIDS}{AIDS}{anomaly-based IDS}
\newacronym{NIDS}{NIDS}{network intrusion detection system}
\newacronym{HIDS}{HIDS}{host-based IDS}
\newacronym{1D}{1D}{one dimensional}
\newacronym{RF}{RF}{radio frequency}
\newacronym{DNN}{DNN}{deep neural networks}
\newacronym{ViT}{ViT}{vision transformer}
\newacronym{CV}{CV}{computer vision}
\newacronym{IoFT}{IoFT}{Internet of Flying Things}